\documentclass{ws-ijmpb}
\def\sign#1{\hbox{\rm \,sign}(#1)}
\begin{document}

\markboth{Roberto D'Autilia}
{The prosody and the music of the human speech}

%
\catchline{}{}{}{}{}
%

\title{PROSODY AND THE MUSIC OF THE HUMAN SPEECH
\footnote{This paper is dedicated to Francesco Guerra.}}

\author{ROBERTO D'AUTILIA}

\address{Dipartimento di Fisica Universit\`a degli Studi di Roma ``La Sapienza''\\
P.le Aldo Moro 2, 00185, Roma, Italy\\
roberto.dautilia@roma1.infn.it}

\author{}

\address{}

\maketitle


\begin{abstract}
We propose the use of a self-oscillating dynamical system --the pre-Galileian clock equation-- for modeling the laryngeal 
tone.
The parameters are shown to be the minimal control needed for generating the prosody of the human speech.
Based on this model, we outline a peak delay detection algorithm for extracting the prosody of the real speech.
\end{abstract}


\section{Introduction}

One of the distinctive features of the human brain is its aptitude for communicating thoughts by speaking.
In spite of its relevance  for exchanging information and emotions, it is very hard to deal with the system  
made by the auditory-phonatory apparatus by means of physical models.
The emotional component of verbal communication relies on the fundamental sound of speech --the laryngeal sound-- 
which is generated  by the vocal cords without any  movement of mouth or tongue.

The glottal sound has been studied for centuries in the framework of different disciplines, and numerous explanations for its unique capability of trasmitting emotions had been proposed.
Among these ideas, one of the most captivating is Jean-Jacques Rousseau theory about the relation between prosody 
and music \cite{Rousseau}.
In his  {\sl Essai sur l'origine del languages, Chapter XII}, he writes:\\
{\it "La col\`ere arrache des cris mena\c cans, que la langue et le palais articulent: mais la voix de la 
tendresse est plus douce, c'est la glotte qui la modifie, et cette voix devient un son; seulement les accens 
en sont plus fr\'equens ou plus rares, les inflexions plus ou moins aigu\"es, selon le sentiment qui s'y joint"}.\\
Later, in the same book, he writes:\\ 
{\it "Qu'on fasse la m\^eme question sur la m\'elodie, la r\'eponse vient d'elle-m\^eme: elle est d'avance dans 
l'esprit des lecteurs. La m\'elodie, en imitant les inflexions de la voix, exprime les plaintes, les cris de 
doleur ou de joie, les menaces, les g\'emissement; tous les signes vocaux des passion sont de son ressort. 
Elle imite les accens des langues, et les tours affect\'es dans chaque idi\^ome \`a certains mouvemens de l'ame: elle n'imite pas seulement, elle parle; et son langage inarticul\'e, mais vif, 
ardent, passionn\'e a cent fois plus d'\'energie que la parole m\^eme"}.

We want to study the relation suggested by Rousseau between musical melodies and corresponding prosodic patterns, 
using a simple model for glottal motion.
The laryngeal sound is generated  by the cyclic motion of  opening and the closing of vocal cords.
At the beginning of the cycle air is pushed by the diaphragm, the vocal cords are drawn together and 
air pressure increase, but when the pressure reaches a critical value it blows the vocal cords apart and 
flows between them. 
Then the vocal cords are then drawn together as a result of the Bernoulli effect \cite{Step}.

To understand this oscillating behavior ``without spring'', we study a simple self-oscillating 
model for the pre-Galileian clock \cite{Andronov} which produces a realistic laryngeal tone.
The model can laso be used as a powerful tool for the analysis of glottal sounds.
The results of the analysis can be compared with some adiastematic notation to suggest a formal correspondence 
between prosody and music.
In this direction it is possible to suggest that the prosody is the drift of the musical gusto evolution, to 
answer to the main question about the nature of stochastic processes which produce ``beautiful'' or at least 
meaningful sequences of sounds \cite{Baffioni}. 

The paper is organized as follows. In the next section we introduce a non-linear dynamical system which exhibit all
the main features of the glottal cycle. In section 3 the parameters of the system are used as time dependent controls
for the glottis, and in the following we present an algorithm for analyzing the control of the recorded sounds.
The form of this control suggest also a delay-line-like behavior for the cochlear apparatus.

\section{The cycle of the glottis.}

The laryngeal tone is the oscillatory variation in air pressure generated by the cyclic movement of the vocal cords. 
At the beginning of each cycle \cite{Cole} the vocal cords are held together by the action of the arytenoid cartilages.
Air is forced into the trachea and when the pressure exceeds a threshold (the value of which depends on the  strength of the vocal cords), 
it opens the vocal cords and flows through the glottis.
Inside the constricted laryngeal passage air pressure falls (its velocity increase) giving rise,  for the Bernoulli principle, 
to the pressure drop closing the vocal cords and completing the cycle. 
The cycle repeats at rates of 130-220 times per second.
The ear perceives the variation in the cycle period as changes in the pitch.

The valve-like behavior producing the laryngeal tone is characteristic of self-oscillating systems.
A self-oscillating system is an apparatus which {\it produces a periodic process at the expense of a non-periodic source of energy}.
Self-oscillations do not depend on the initial condition but are determined by the properties of the system itself.
Examples of self-oscillating systems include the electric bell, saw-tooth signal generators as well as wind and string musical instruments \cite{Andronov}.

Several realistic models for the glottal behavior have been proposed over the years starting with the celebrated 
``two-mass model''~\cite{Flan,Ish,Titzea,Titzeb}.
We will now present a minimal mechanism thet exibits all the main features of glottal behavior.
In particular we want to make explicit both the dependence of the oscillation period on the forces acting on the system, and the features of the 
trigger mechanism producing the self-oscillations.

Let  $s\in[-1,1]$ the variable related to the aperture of the glottis: in the extreme positions  $s=-1$ indicates that the glottis is completely closed 
and $s=1$ that the vocal cords are open.
We represent $s$ in the $[-1,1]$ interval in agreement with the usual representation of the acoustic signals.
Assuming that the laringeal tone is proportional to the opening, $s$ can be assimilated to the signal itself
\begin{equation}
\label{segnale}
[T_0, T_1]\ni t\rightarrow s(t)\in[-1,1]
\end{equation}
where $[T_0,T_1]$ is a time interval.
For simplicity we assume that the forces act on the glottis istantaneously: for
$s(t)\geq+s_0$ the Bernoulli effect produce a force which closes the glottis,  for $s(t)\leq-s_0$ the pressure opens the glottis, and for
$-s_0<s(t)<+s_0$ no force is acting on the glottis.
This approximation is useful for solving the model, but can be easily relaxed in computer simulations.
When the glottis is opening, the force $P(s)$ acting on it is negative and when it is closing $P(s)>0$.
Therefore  over the interval $-s<s<+s$ the force  $P=P(s)$ is a twovalued function of the variable $s$ representing
the opening of the glottis.
Following the Andronov argument \cite{Andronov} $P(s)$ imposes limitations on the shape of the phase plane trajectories,
since assigning  $(s, \dot s)$ does not uniquely determines the state of the system  where $-s_0<s<s_0$ .
Instead we have to use a phase surface with two half-planes superimposed: (a) $s<s_0$ and (b) $s>-s_0$.
The points on this two-sheet phase surface have a one to one correspondence with the states of the system, 
the passage of the 
representative point from sheet (a) to the sheet (b) occurs for $s=+s_0$, the reverse passage for $s=-s_0$, 
and the abscissa remains unvaried in both the cases.

To further semplify the model we assume that the force $P(s)$ applied to the glottis by the air pressure is
constant in absolute value: $P(s)=+P_0$ for the closing and $P(s)=-P_0$ for the opening.
To model the vocal cords tension we introduce the constant resistence $f_0$, which does not depend on 
the position of the glottis.
On the basis of these simple assumptions it is possible to describe two different laringeal sounds.
The first one, that we call prosody, does not have a natural period because there is no elastic force contributing to 
the movement of the vocal cords.
This system does not exhibit high stability and is therefore a good model for cases in which the period of oscillation 
has to be sensitive to variations in control parameters.
The second type has a natural period due to the elastic term, and  in absence of feeding can perform damped
oscillations.
This model can be used to describe singing, but is out of the scope of this paper.

The dynamic equation for the model without the elastic force is
\begin{equation}
\label{moto}
m\ddot s=f(s,\dot s)+P(s)
\end{equation}
where $m$ is the glottis mass, $P=P(s)$ the force produced by air pressure on it
and $f(s,\dot s)$ is the resistence of the vocal cords.
If we assume $f(s,\dot s)=-f_0\sign{\dot s}$ during the motion ($\dot s\neq0$), equation (\ref{moto}) becomes
\begin{equation}
\label{esplic}
m\ddot s=-f_0\sign{\dot s}\pm P_0
\end{equation}
Introducing the variables $x=s/s_0$ and $z=+\sqrt{P_0/ms_0}t$, equation~(\ref{esplic}) can be rewritten as
\begin{equation}
\label{due}
\cases{\dot x=y\cr
\dot y=-F \sign y-(-1)^n}
\end{equation} 
where the dot operator is now the derivative with respect to the scaled time $z$, $y\neq0$, $F=f_0/P_0$ and $n$ is 
even when the glottis is closing and odd when the glottis is opening.
If the air pressure outside the glottis is a linear funcion of the glottis aperture, the $x$
variable can be studied as the waveform of the laryngeal tone.

If $F\geq1$ and the glottis is at rest ($y=0$) the force produced by the air pressure can not open the vocal cords,
$\dot y=0$ and the point $(x,0)$ is the equilibrium state.
Therefore, for the case when  $f_0<P_0$ the system has no states of equilibrium.
From equation~(\ref{due}) we have for the (a) half plane ($x<+1$ and $P=-P_0$)
\begin{equation}
\frac{dy}{dx}=\frac{1-F \sign y}{y}
\end{equation}
and integrating we have $y^2/2-(1+F)x=k$ for $y<0$ and $y^2/2-(1-F)x=k$ for $y>0$, $k=const$, so that the phase path on the
sheet (a) is made by two parabolae and the representative point moves to the left on the lower half of the sheet
($y<0$) and to the right on the upper one.
All the phase paths on sheet (a) reach their boundaries on the semiaxis $x=+1$, $y>0$, and the phase paths on 
sheet (b) are symmetrical with respect to the origin of the coordinates.

Following the Andronov treatment \cite{Andronov} we draw the two axes $v$ where $x=-1$, $y=-v<0$ and $v'$ where 
$x=1$, $y=v'>0$ and consider the sequence of the points of intersections with them of an arbitrary phase path: 
$v, v_1, v_2, v_3, \ldots$.
The representative point pass at the point $(-1,-v)$ from sheet (b) to sheet (a) and reach the axis of
the abscissa at the point $(-\xi,0)$, where $\xi$ is given by the equation $v^2=2(1+F)(\xi-1)$.
Then the representative point moves on the upper half of the sheet (a) and reaches its boundary at $(+1,v_1)$ where
$v_1$ is given by $v_1^2=2(1-F)(\xi+1)$.
The phase path on (a) establishes a one-to-one correspondence between the points of the axes $v$ and $v'$, a point 
transformation where the sequence function is parametrized by the peak $\xi$ of the laryngeal tone.
Then the representative point passes on the sheet (b) and reaches the semi-axis $v$
on $(-1, -v_2)$, where $v_2$ is determined by the same sequence (due to the symmetry of the sheets (a) and (b)) and the point
transformation of $v'$ in $v$ is the same of $v$ in $v'$.
The fixed point $v=v_1=\bar v$, corresponding to a symmetric limit cycle, is given by the equation
$(1+F)(\bar\xi-1)=(1-F)(\bar\xi+1)$ and is determined by the glottal peak 
\begin{equation}
\label{ampiezza}
\bar\xi=\frac{1}{F}
\end{equation}
and $\bar v^2=2\frac{1-F^2}{F}$.
The Lamerey's ladder given by the two curves $v^2=v^2(\xi)$ and $v_1^2=v_1^2(\xi)$, tends to the fixed point
if $v^2=2(1+F)(\xi-1)$ has a steeper slope than $v_1^2=2(1-F)(\xi+1)$ and the unique periodic motion is reached
from any initial conditions.
The period of the oscillation can be easily computed by the evaluation of the transit time on the parabolae arcs
\begin{equation}
\label{periodo}
T=4\sqrt{\frac{2s_0m}{P_0}}\frac{1}{\sqrt{F(1-F^2)}}
\end{equation}
and depends on both the air pressure and the vocal cord resistence.
\begin{figure}
\centerline{\psfig{file=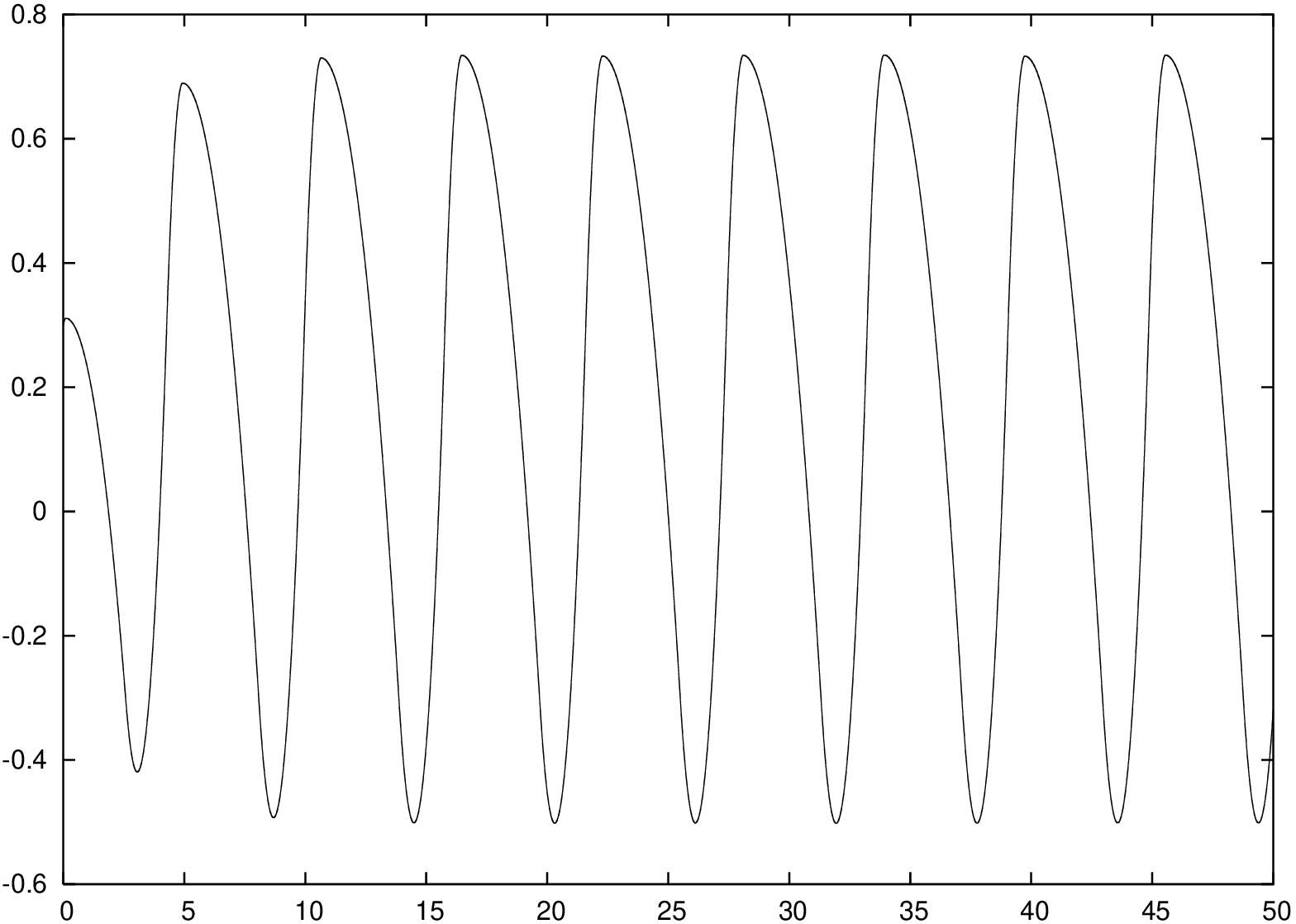,width=4cm} \psfig{file=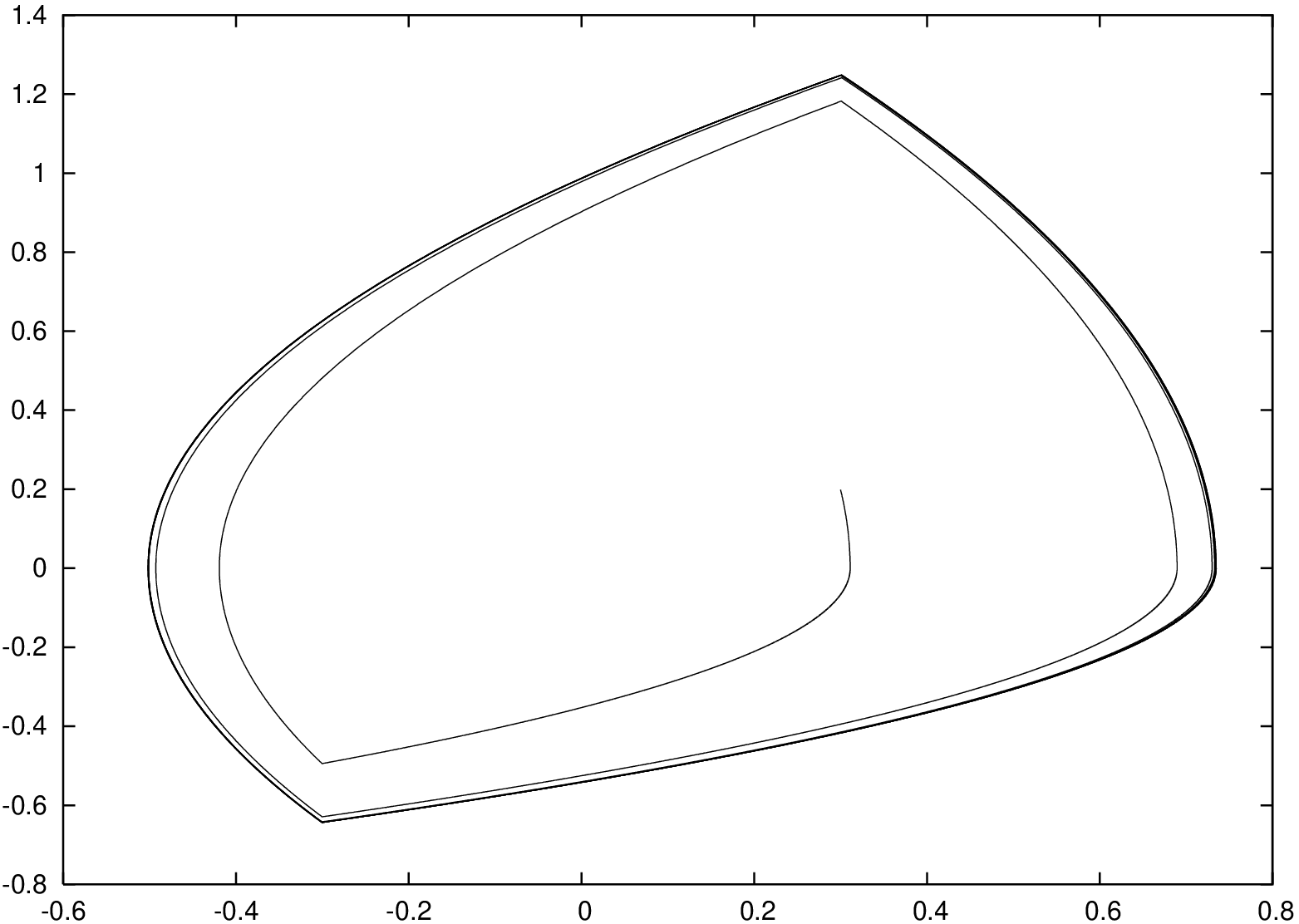,width=4cm}}
\vspace{-50pt}
\caption{The plot of the signal as a function of time (on the left) and the corresponding phase path (on the right), 
where the two sheets are glued together and the vocal cords resistence is  different for the opening and the closing.
The value of $s$ has been scaled to fit in the $[-1,1]$ interval.}
\label{figuraa}
\end{figure}

If we assume that the vocal cords resistence is not the same for the opening and closing glottis, 
the equation~(\ref{due}) gives the signal of figure~\ref{figuraa}. The corresponding sound is very similar to 
the sound obtained by placing a microphone on the throat and directly recording the glottal sound. 

\section{The prosodic patterns of the controlled glottis}
The stability of equations~(\ref{esplic}) can be evaluated by the ratios of the percentage variation of the period to the
percentage variation of the two parameters, and shows to be very small \cite{Andronov}.
Because of this low stability, the system is not suitable for building clocks.
However its sensitivity to variation in the parameters and the fast convergence to the limit cycle, 
make it a suitable model for a controlled glottis.
 
Let us suppose that the air pressure and the strength of the vocal cords are controlled by the speaker.
Controlling $P(t)$ and $f(t)$ it is possible to change the amplitude (\ref{ampiezza}) and the period (\ref{periodo}) 
of the laryngeal tone to produce the prosodic patterns conveying the non semantic aspect of the speech.

In order to decode the prosodic pattern we neet to know the time evolution of the control function
\begin{equation}
\label{controllo}
C(t)=\Big(f\big(\bar\xi(t),T(t)\big),P\big(\bar\xi(t),T(t)\big)\Big)
\end{equation}
In other wordss we suggest that when the cochlea decode the prosody of the speech it needs only to detect the 
value of $\bar\xi$ and the time delay of $\bar\xi$ with respect to the previous one.
This stands in agreement with recent studies about the delay lines inside the cochlea~\cite{Mammano}.
Therefore the system can be controlled by a piecewise constant function which changes the period and the amplitude.
\begin{figure}
\centerline{\psfig{file=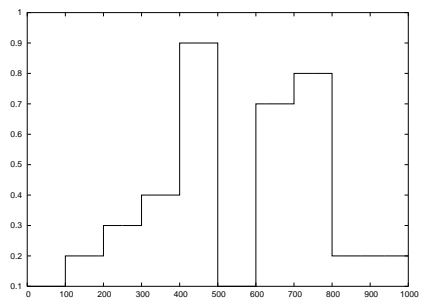,width=4cm}}
\vspace{-50pt}
\caption{A piecewise constant $F$ control function, determining the prosodic pattern.}
\label{figurab}
\end{figure}
If we choose a control function (see figure~\ref{figurab}) we determine the time evolution of $s(t)$ given by
equation~(\ref{esplic}).
The resulting phase path and waveform are plotted in figure~\ref{figurac}.
Note that in this case the phase paths intersect 
because now the system is controlled and non-autonomous. 
\begin{figure}
\centerline{\psfig{file=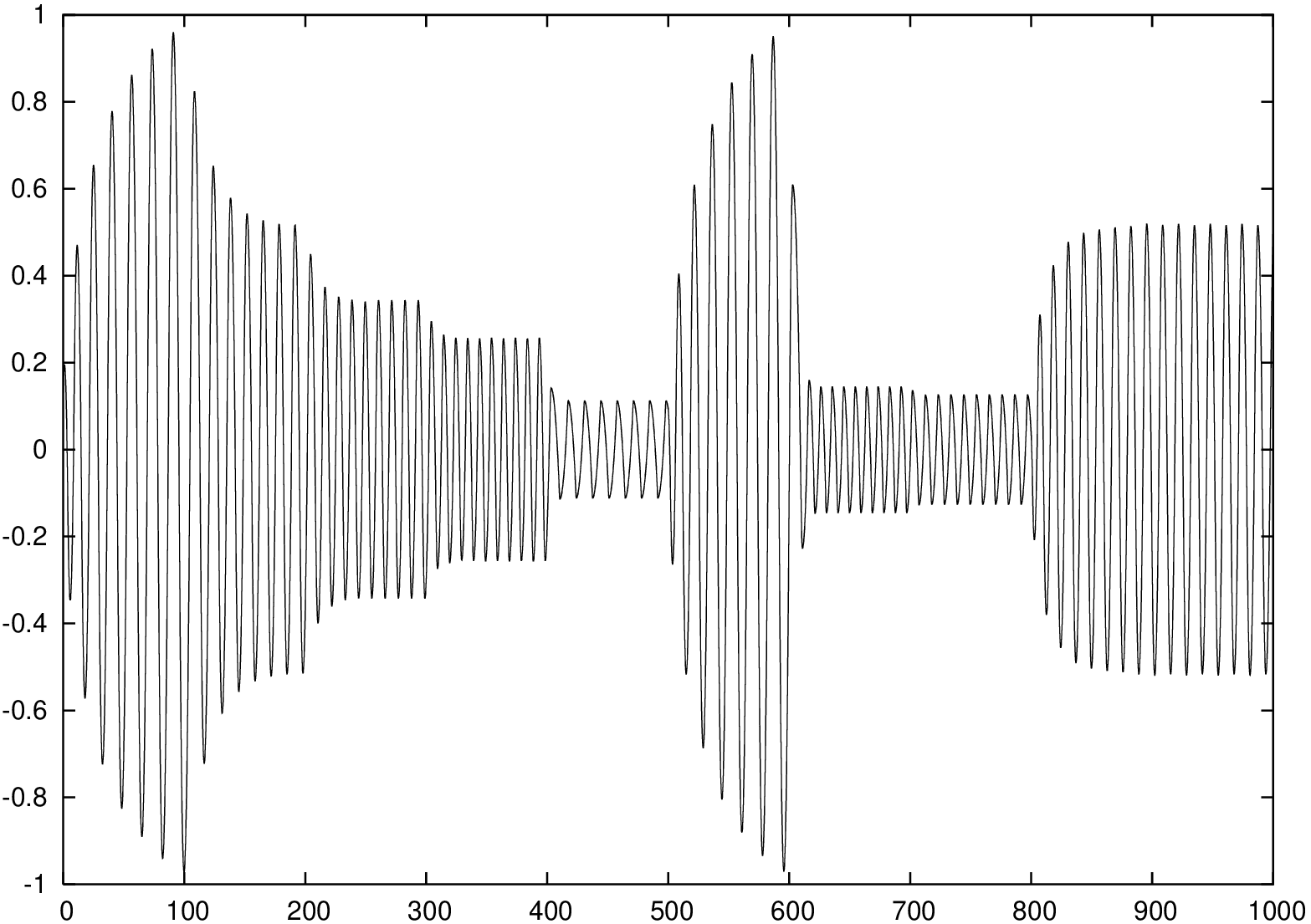,width=4cm}\psfig{file=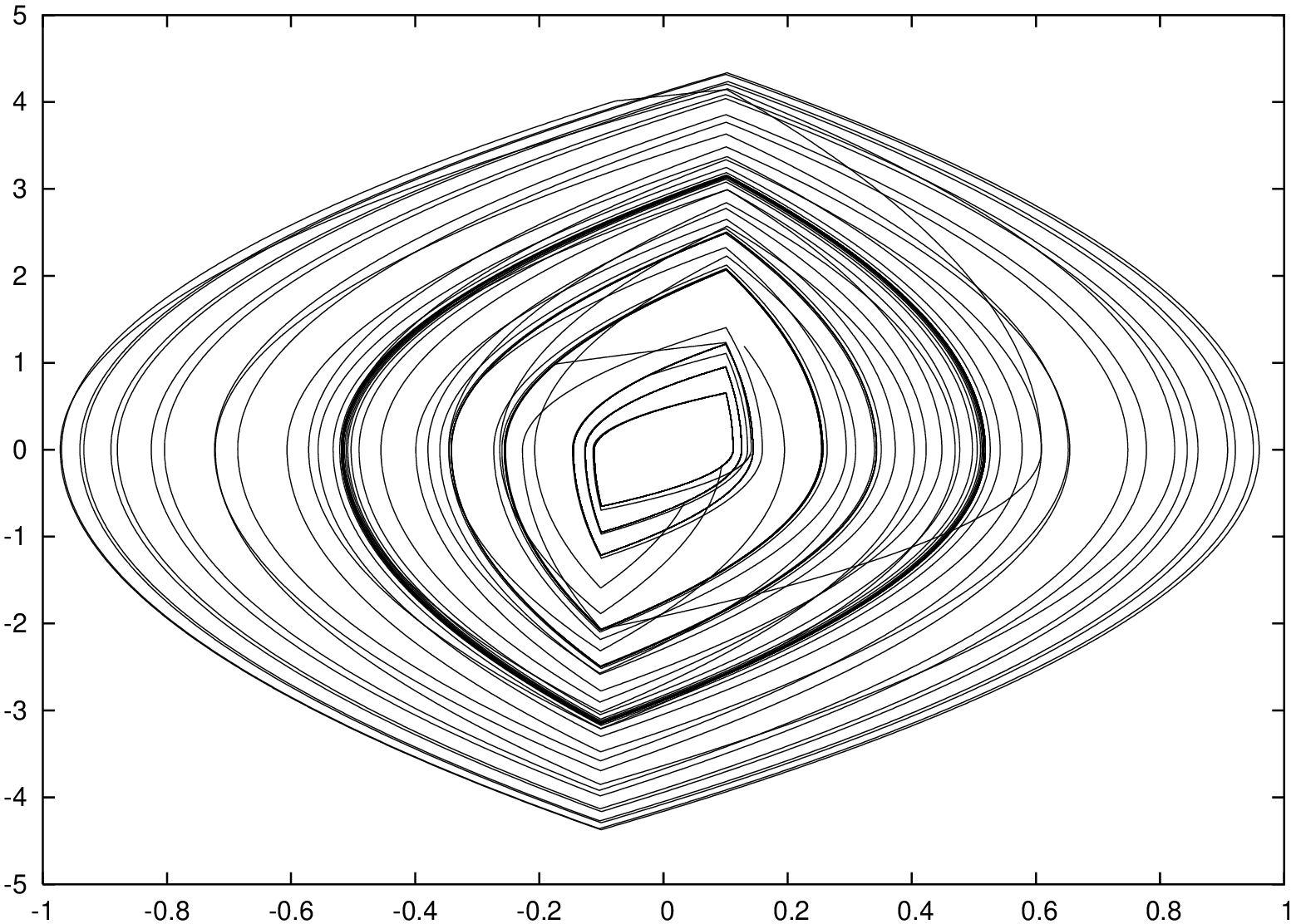,width=4cm} }
\vspace{-50pt}
\caption{The plot of the glottal sound controlled by the signal of figure~\ref{figurab} (left) and the corresponding
 phase path (right).}
\label{figurac}
\end{figure}

We observe also that the characteristic time of the control
change is larger than the period of the wave, in agreement with studies comparing of the laryngeal
tone period with the muscular control time \cite{Step}. 
Therefore we state that the prosodic information is all contained in $C$ and that in order to
decode the prosodic pattern the listener needs to decode the $C$ function.

\section{Perception of the laryngeal tone}
The dynamical system given by equation~(\ref{due}) can be used to detect and represent the prosody of real speech.
To accomplish this we only need to know the discrete function
\begin{equation}
\label{picchi}
[T_0, T_1]\ni t\rightarrow \bar\xi(t)\in[0,1]
\end{equation}
Even though the sound of (\ref{picchi}) is different from the fundamental sound (\ref{segnale}), their prosody is 
identical.
Moreover during the speech process the time interval between two consecutive $\bar\xi$ is changed 
countinuously to produce the prosodic meaning.

\begin{figure}
\centerline{\psfig{file=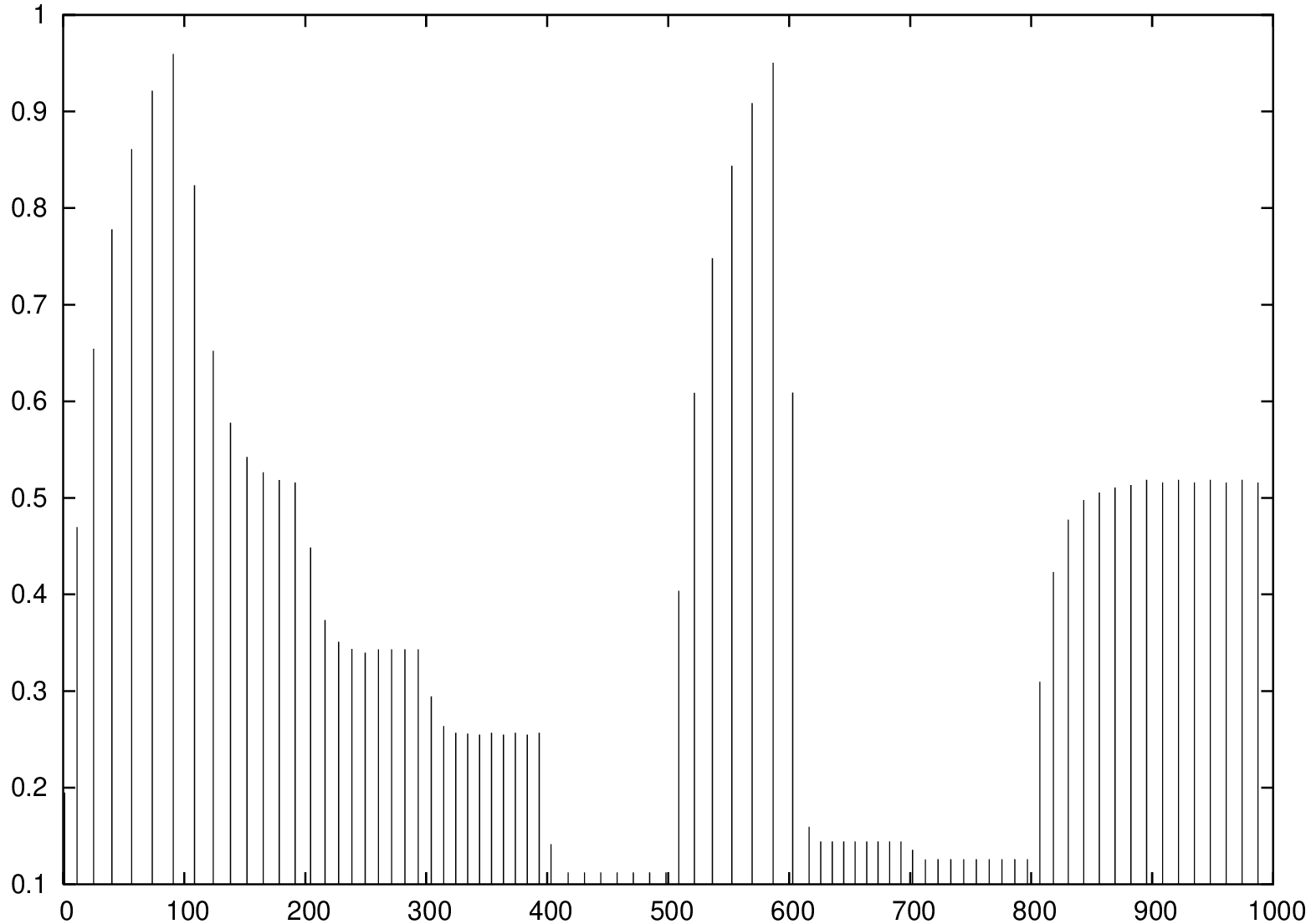,width=4cm}\psfig{file=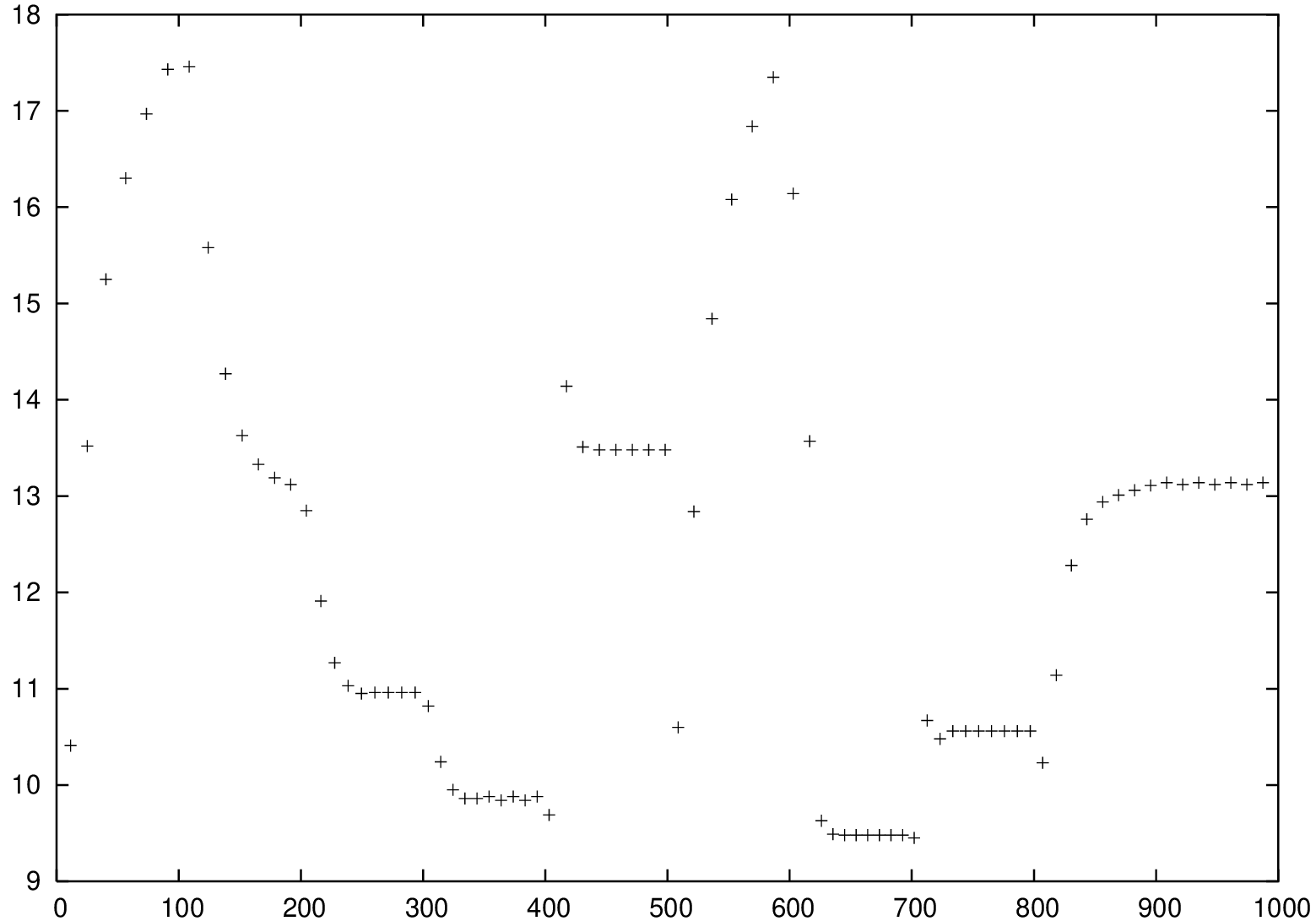,width=4cm} }
\vspace{-50pt}
\caption{The glottal peaks for the wave in fig figure~\ref{figurac} (left) and the $\bar\xi$ delay (right) as function of
the time.}
\label{figurad}
\end{figure}

The $\bar\xi(t)$ function can be extracted from a recorded sound by means of the simple procedure of sliding a
temporal fixed window on the signal, and selecting for each window the value of $s(t)$ not 
exceeded backward and forward  \cite{Dau}.
The resulting signal is shown in figure~\ref{figurad} on the left.

In order to obtain the control $C$ function we have to know the $T(t)$ value also. 
To do this we plot the delay of each peak $\bar\xi(\tau_i)$ with respect to the previous one $\bar\xi(\tau_{i-1})$ 
as funtion of time (figure~\ref{figurad} right) where $\tau_1, \tau_2, \ldots, \tau_N$ are the times where $\bar\xi(t)$ 
is non zero.

In this way it is possible to construct a map of the control $C$ and therefore to decode the prosody of the speech.
The amplitude and the delay of $\bar\xi$ are the only information needed. 
The control function represented in figure~\ref{figurab} can be obtained by the data represented in figure~\ref{figurad} via
the equations~(\ref{ampiezza}) and~(\ref{periodo}). If we take the fundamental given by the Fast Fourier Transform 
of $s(t)$ we obtain a rough map of $C$ because the FFT needs some cycles to detect the period. 

When applied to the real speech, the ``peak detection'' yelds maps which resemble the neumatic notation of the gregorian
chant. This is not surprising given that during the middle age the rules of the correspondence between text 
and the music where very strict, and the composer had to put the known melodies together rather then inventing 
anything new ones \cite{Idel}. 
Following the ideas of Rousseau, the algorithm proposed for the analysis and the synthesis of the laringeal tone
can be a tool for studying the correspondence between prosody and music by means of the comparison of the contol maps 
with the adiastematic notation system \cite{Dau}. 

\section{Conclusion}

Although it is still not clear how the proposed  system is related to the fluidodynamics of the glottis, the 
self-oscillating system used to model the laryngeal behavior is very simple and exhibits good agreement with the
main features of the glottal sounds. 
It produces a realistic laryngeal sound an can be used to extract the prosody from a recorded sound.
Its main features, shared with many ``natural'' oscillators is the lacking of the elastic term.
The control of this system suggests also that the cochlear behavior is related to delay line detection.
The equation is very simple to simulate in real time on a computer, and can be used for generating 
prosodic patterns or for manipulating the prosodic meaning of recorded sentencces. 

\section*{Acknowledgements}

I wish to thank Francesco Guerra for the helpful discussions and the scientific and human support to this research, and
Assaf Talmudi for the carefully reading of this manuscript.
%
%
%
%

\end{document}